\documentclass[aps,pra,reqno,twocolumn,floatfix,superscriptaddress,nofootinbib]{revtex4-1}

\usepackage{amsmath}
\usepackage{amsfonts}
\usepackage{amssymb}
\usepackage{graphicx}
\usepackage{color}
\usepackage{enumitem}

\usepackage{hyperref}

\newcommand{\ket}[1]{\left|#1\right\rangle}
\newcommand{\bra}[1]{\left\langle #1\right|}

\begin{document}
\title{Quantum metrology including state preparation and readout times}
\author{Shane Dooley}
\email[]{dooleysh@nii.ac.jp}
\affiliation{National Institute of Informatics, 2-1-2 Hitotsubashi, Chiyoda-ku, Tokyo 101-8430, Japan.}

\author{William J. Munro}
\affiliation{NTT Basic Research Laboratories, 3-1 Morinosato-Wakamiya, Atsugi, Kanagawa 243-0198, Japan.}
\affiliation{National Institute of Informatics, 2-1-2 Hitotsubashi, Chiyoda-ku, Tokyo 101-8430, Japan.}

\author{Kae Nemoto}
\affiliation{National Institute of Informatics, 2-1-2 Hitotsubashi, Chiyoda-ku, Tokyo 101-8430, Japan.}
\date{\today}

\begin{abstract}
There is growing belief that the next decade will see the emergence of sensing devices based on the laws of quantum physics that outperform some of our current sensing devices. For example, in frequency estimation, using a probe prepared in an entangled state can, in principle, lead to a precision gain compared to a probe prepared in a separable state. Even in the presence of some forms of decoherence, it has been shown that the precision gain can increase with the number of probe particles $N$. Usually, however, the entangled and separable state preparation and readout times are assumed to be negligible. We find that a probe in a maximally entangled (GHZ) state can give an advantage over a separable state only if the entangled state preparation and readout times are lower than a certain threshold. When the probe system suffers dephasing, this threshold is much lower (and more difficult to attain) than it is for an isolated probe. Further, we find that in realistic situations the maximally entangled probe gives a precision advantage only up to some finite number of probe particles $N_\text{cutoff}$ that is lower for a dephasing probe than it is for an isolated probe.
\end{abstract}

\maketitle

\section{Introduction}

In quantum metrology the goal is often to estimate some unknown parameter $\omega$ by measuring a probe system whose quantum state $\hat{\rho}_\omega$ depends on that parameter \cite{Hel-69, Par-09}. Usually, the dependence comes about through some $\omega$-dependent dynamics. It is well known that under ideal conditions, entanglement in the probe can be exploited to increase the precision of the estimate \cite{Win-92, Gio-04, Gio-06}. To determine the extent of the precision gain in practice, it is important to consider realistic, non-ideal conditions. Here we investigate the effect of including measurement and readout times in quantum metrology.

\begin{figure}[b]
    \centering
    \includegraphics[width=85mm]{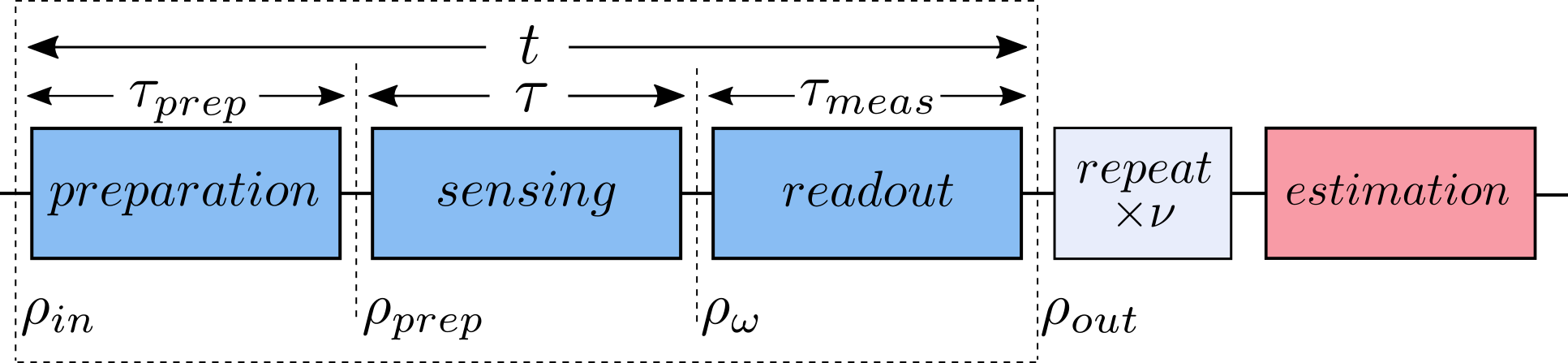} 
    \caption{Illustration of the main steps of a quantum metrology protocol. The preparation, sensing and readout steps may be repeated $\nu$ times before the final estimation step. Usually the state preparation time $\tau_\text{prep}$ and the readout time $\tau_\text{meas}$ are assumed to be negligible. In practice, however, they are always finite.} \label{fig:schematic}
\end{figure}

A quantum metrology protocol has four main steps: ($i$) [\emph{preparation}] the probe system is initialised; $(ii)$ [\emph{sensing}] the probe system evolves in time, picking up a dependence on the unknown parameter $\omega$; $(iii)$ [\emph{readout}] the probe system is measured to extract the information about the parameter $\omega$; $(iv)$ [\emph{estimation}] the parameter $\omega$ is estimated based on the measurement results. Steps $(i)$-$(iii)$ may be repeated $\nu$ times before the final processing of the measurement results in the estimation step. If each repeat of steps $(i)$-$(iii)$ takes time $t$ to complete, the total time for the protocol is $T = \nu t$. The protocol is illustrated in Fig. \ref{fig:schematic}. The error $\delta\omega$ of the final estimate depends on the state that is prepared in step $(i)$. For example, in the case of frequency estimation using a probe consisting of $N$ two-level systems, a separable state can give -- at best -- a sensivity that scales as $\delta\omega \sim 1 / \sqrt{N}$ (standard quantum scaling) while entangled states (such as GHZ states) can, in principle, give sensitivity $\delta\omega \sim 1 / N$ (Heisenberg scaling) \cite{Win-92, Gio-04, Gio-06}. This is a significant improvement when $N$ is a large number. When decoherence is taken into account the precision gain using entangled states is somewhat diminished: with parallel Markovian dephasing we have a return to standard quantum scaling $\delta\omega \sim 1/\sqrt{N}$ \cite{Hue-97, Esc-11, Dem-12}, but with parallel non-Markovian dephasing we can achieve $\delta\omega \sim 1/N^{3/4}$ \cite{Mat-11, Chi-12}, and  with transverse Markovian dephasing the scaling $\delta\omega \sim 1/N^{5/6}$ is possible with entangled probe states \cite{Cha-13}. These results suggest that we should use the preparation stage to generate an entangled state that is sensitive to the $\omega$-dependent dynamics in the sensing period. However, we note that it is usually assumed that the preparation and readout steps take a negligible amount of time. 

The error $\delta\omega$ also typically decreases with the amount of time given to the sensing stage \cite{Yua-15}, which we denote $\tau$. This suggests that we should maximise the sensing time. However, when the preparation stage takes a non-neglibible amount of time this leads to a tradeoff: we can use the preparation stage to generate a state that is sensitive to small changes in the unknown parameter, but the time taken to prepare this state is then unavailable for sensing. In practice, the situation is further complicated by the fact that the time $t$ available for each prepare-sense-readout round is always limited by decoherence of the probe system. Here we investigate this tradeoff via the prototypical example of frequency estimation. 

\section{Model} 

A probe consisting of $N$ two-level systems is initialised in a state $\hat{\rho}_\text{prep}$, which requires a preparation time $\tau_\text{prep}$. This state then evolves for a sensing time $\tau$ by the Hamiltonian $\hat{H}_{\omega} = \frac{\hbar\omega}{2}\sum_{i=1}^N \hat{\sigma}_z^{(i)}$ where $\omega$ is the parameter to be estimated. During the sensing period each particle in the probe also interacts with a bath leading to dephasing of the probe state by the quantum channel $\Lambda^{\otimes N}$ where for the $i$'th particle the action of the channel is $\Lambda [\hat{\rho}^{(i)}] = \frac{1 + e^{-\Gamma (\tau)}}{2} \hat{\rho}^{(i)} + \frac{1 - e^{-\Gamma (\tau)}}{2}\hat{\sigma}_z^{(i)} \hat{\rho}^{(i)} \hat{\sigma}_z^{(i)}$. Here $\Gamma (\tau)$ is a real function of the sensing time that is determined by the details of the bath (see Appendix \ref{app:bath}). We can think of the dephasing channel $\Lambda^{\otimes N}$ and the unitary evolution operator $\hat{U}_\omega = \exp[-i\tau\hat{H}_\omega /\hbar]$ as operating simultaneously during the sensing period, since the dynamics due to $\Lambda^{\otimes N}$ and $\hat{U}_\omega$ commute with each other. After the sensing period has ended, the probe is in the $\omega$-dependent state $\rho_\omega = \hat{U}_\omega \Lambda^{\otimes N}[\hat{\rho}_\text{prep}] \hat{U}_\omega^{\dagger} $. The final readout takes a time $\tau_\text{meas}$ so that the total time for a single prepare-sense-readout round is $t = \tau_\text{prep} + \tau + \tau_\text{meas}$. This is repeated a number of times $\nu \gg 1$, so that the total experiment time is $T = \nu t$. Denoting $\tilde{\tau} = \tau_\text{prep} + \tau_\text{meas}$ for convenience, we have $t = \tilde{\tau} + \tau$. 

The error in the estimate of $\omega$ is bounded by the quantum Cramer-Rao inequality \cite{Bra-94}: \begin{equation} \delta\omega \geq \frac{1}{\sqrt{\nu \mathcal{F}(\hat{\rho}_{\omega},\tau) } } = \frac{1}{\sqrt{T\mathcal{F}(\hat{\rho}_{\omega},\tau) / (\tilde{\tau} + \tau) } } , \label{eq:CR} \end{equation} where \begin{equation} \mathcal{F}(\hat{\rho}_{\omega},\tau) = 2 \sum_{i,j} \frac{1}{\lambda_i + \lambda_j} \left| \bra{\phi_j}\frac{d \hat{\rho}_\omega}{d\omega} \ket{\phi_i} \right|^2 , \label{eq:qfi} \end{equation} is the quantum Fisher information of the state $\hat{\rho}_\omega$ with eigenvalues $\{ \lambda_i \}$ and eigenstates $\{ \ket{\phi_i} \}$. We have explicitly written the dependence on the sensing time $\tau$ in the argument of $\mathcal{F}$. Maximising the quantity $\mathcal{F}(\hat{\rho}_\omega,\tau)/(\tilde{\tau}+\tau)$ on the right hand side of Eq. \ref{eq:CR} over the sensing time $\tau$ gives the optimum precision: \begin{eqnarray} \delta\omega \geq \delta\omega_\text{opt} &=& \frac{1}{\sqrt{T \max_\tau [ \mathcal{F}(\hat{\rho}_\omega,\tau)/(\tilde{\tau}+\tau) ] }} \nonumber  \\ &=& \frac{1}{\sqrt{T \left[ \mathcal{F}(\hat{\rho}_{\omega},\tau_\text{opt})/(\tilde{\tau}+\tau_\text{opt}) \right]}} , \label{eq:prec_limit} \end{eqnarray} for a given prepared state $\hat{\rho}_\text{prep}$, where $\tau_\text{opt}$ is the optimal sensing time. We note that by using the quantum Fisher information we implicitly assume that the optimal POVM can be implemented at the readout stage. 

We would like to compare $\delta\omega_\text{opt}$ for two different choices of the prepared state $\hat{\rho}_\text{prep}$: \begin{enumerate}[label=(\roman*)] \item the separable state $\ket{\psi^\text{sep}} = \left[ \frac{1}{\sqrt{2}} ( \ket{0} + \ket{1} ) \right]^{\otimes N}$, \item the maximally entangled GHZ state $\ket{\psi^\text{ent}} = \frac{1}{\sqrt{2}} ( \ket{0}^{\otimes N} + \ket{1}^{\otimes N} )$. \end{enumerate} The states $\ket{0}$ and $\ket{1}$ here are the eigenstates of $\hat{\sigma}_z$ for each two-level system. For the separable state ($\hat{\rho}_\text{prep}^\text{sep} = \ket{\psi^\text{sep}}\bra{\psi^\text{sep}}$) we denote the preparation and readout time, the sensing time, and the optimal precision as $\tilde{\tau} = \tilde{\tau}^\text{sep}$, $\tau_\text{opt} = \tau_\text{opt}^\text{sep}$ and $\delta\omega_\text{opt} = \delta\omega_\text{opt}^\text{sep}$ respectively (i.e., with the superscript ``sep''). This notation allows us to distinguish these from the corresponding quantities $\tilde{\tau} = \tilde{\tau}^\text{ent}$, $\tau_\text{opt} = \tau_\text{opt}^\text{ent}$ and $\delta\omega_\text{opt} = \delta\omega_\text{opt}^\text{ent}$ when the entangled state ($\hat{\rho}_\text{prep}^\text{ent} = \ket{\psi^\text{ent}}\bra{\psi^\text{ent}}$) is prepared. For a fair comparison, we assume that in both cases the physical resources $N$ and $T$ are the same. We note that we have assumed that the state preparation is ideal, that is, the states $\hat{\rho}_\text{prep}^\text{sep} = \ket{\psi^\text{sep}}\bra{\psi^\text{sep}}$ and $\hat{\rho}_\text{prep}^\text{ent} = \ket{\psi^\text{ent}}\bra{\psi^\text{ent}}$ can be prepared with perfect fidelity. The precision that can be achieved with these two prepared states can then be compared with the ratio \cite{Chi-12}: \begin{eqnarray} r &=& \left( \delta\omega_\text{opt}^\text{sep} / \delta\omega_\text{opt}^\text{ent} \right)^2 \nonumber \\ &=& \frac{\max_{\tau} \left[ \mathcal{F}(\hat{\rho}_\omega^\text{ent},\tau)/(\tilde{\tau}^\text{ent} + \tau) \right]}{\max_\tau \left[ \mathcal{F}(\hat{\rho}_\omega^\text{sep},\tau)/(\tilde{\tau}^\text{sep} + \tau) \right]} . \label{eq:r1} \end{eqnarray} If $r\leq 1$ there is no advantage in preparing the maximally entangled GHZ state. If $r>1$ there is an advantage in preparing the entangled GHZ state, even taking the preparation and readout time $\tilde{\tau} = \tau_\text{prep} + \tau_\text{meas}$ into account. 

To derive an explicit expression for $r$ the first step is to calculate the quantities $\mathcal{F}(\hat{\rho}_\omega^\text{sep}, \tau)$ and $\mathcal{F}(\hat{\rho}_\omega^\text{ent}, \tau)$ via Eq. \ref{eq:qfi}. In the separable case, finding the eigenvalues and eigenvectors of the state $\hat{\rho}_\omega^\text{sep} = \hat{U}_\omega \Lambda^{\otimes N} [\rho_\text{prep}^\text{sep}]\hat{U}_\omega^{\dagger}$ is relatively easy since it is a tensor product of $N$ identical two-dimensional mixed states. Using these eigenvalues and eigenvectors in Eq. \ref{eq:qfi} we obtain $\mathcal{F}(\hat{\rho}_\omega^\text{sep}, \tau) = N \tau^2 e^{-2\Gamma(\tau)}$. Similarly, the eigenstates and eigenvalues of $\hat{\rho}_\omega^{\text{ent}} = \hat{U}_\omega \Lambda^{\otimes N} [\rho_\text{prep}^\text{ent}]\hat{U}_\omega^{\dagger}$ are easy to calculate since the state evolves in a two-dimensional subspace (spanned by $\ket{0}^{\otimes N}$ and $\ket{1}^{\otimes N}$) of the whole $2^N$-dimensional state space. The quantum Fisher information in this case is $\mathcal{F}(\hat{\rho}_\omega^\text{ent}, \tau) = N^2 \tau^2 e^{-2N\Gamma(\tau)}$. The next step is to find the optimal sensing times $\tau_\text{opt}^\text{ent}$ and $\tau_\text{opt}^\text{sep}$ that maximise the numerator and denominator of Eq. \ref{eq:r1}, which leads to: \begin{eqnarray} r &=& N \left( \frac{\tilde{\tau}^\text{sep} + \tau_\text{opt}^\text{sep}}{\tilde{\tau}^\text{ent} + \tau_\text{opt}^\text{ent}} \right) \left( \frac{ \tau_\text{opt}^\text{ent} }{\tau_\text{opt}^\text{sep} } \right)^2 \times \nonumber \\ &&\quad \exp\left[ -2 N \Gamma ( \tau_\text{opt}^\text{ent}) + 2 \Gamma( \tau_\text{opt}^\text{sep} ) \right] . \label{eq:general_r} \end{eqnarray} These optimal sensing times depend on the form of the function $\Gamma (\tau)$, which itself depends on the details of the bath. We first consider an isolated probe.

\begin{figure*}[ht]
    \centering
    \href{https://youtu.be/qOZ6UYPyb4M}{\includegraphics[width=59mm]{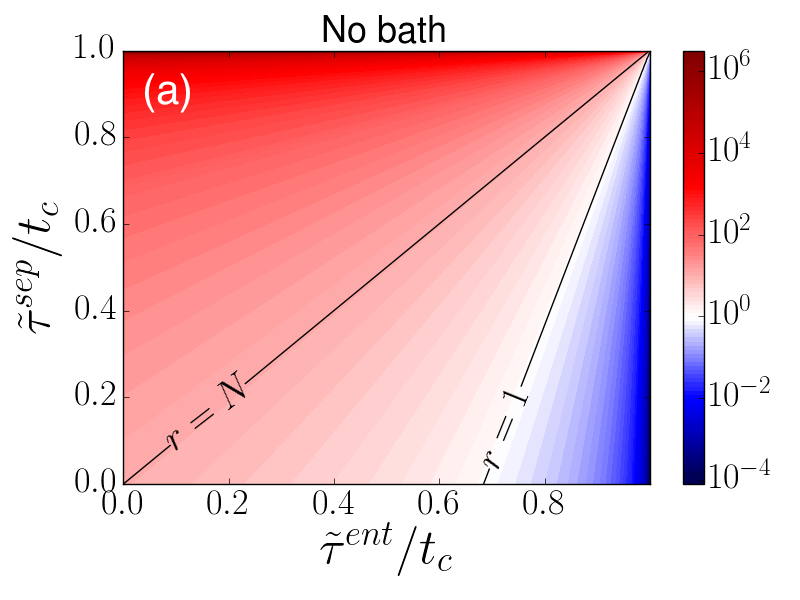}}
    \href{https://youtu.be/kb2KPl3_4K0}{\includegraphics[width=59mm]{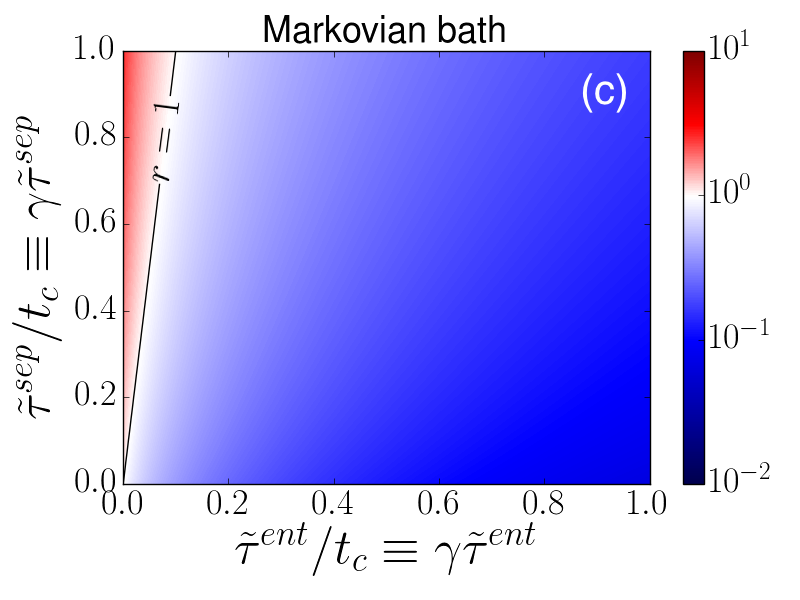}}
    \href{https://youtu.be/aUQYTy_tif4}{\includegraphics[width=59mm]{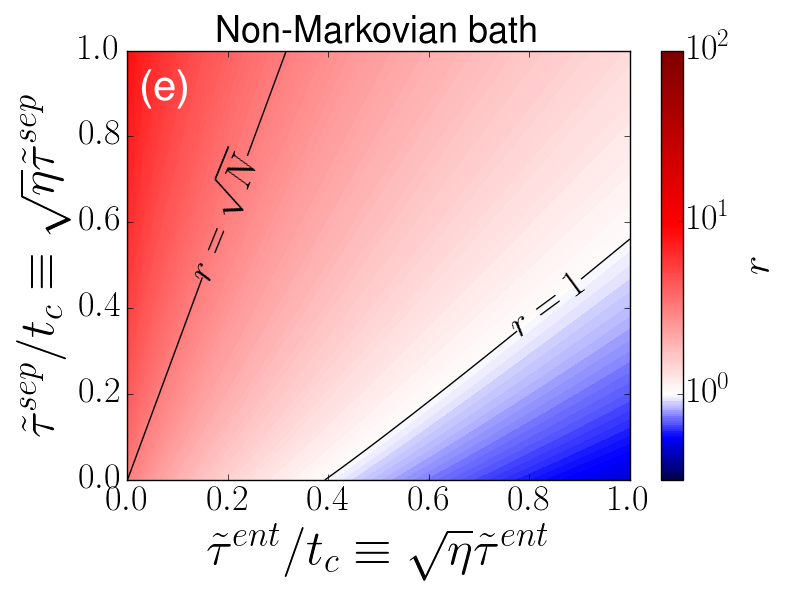}}
    \href{https://youtu.be/wEK8G7Pc2T4}{\includegraphics[width=59mm]{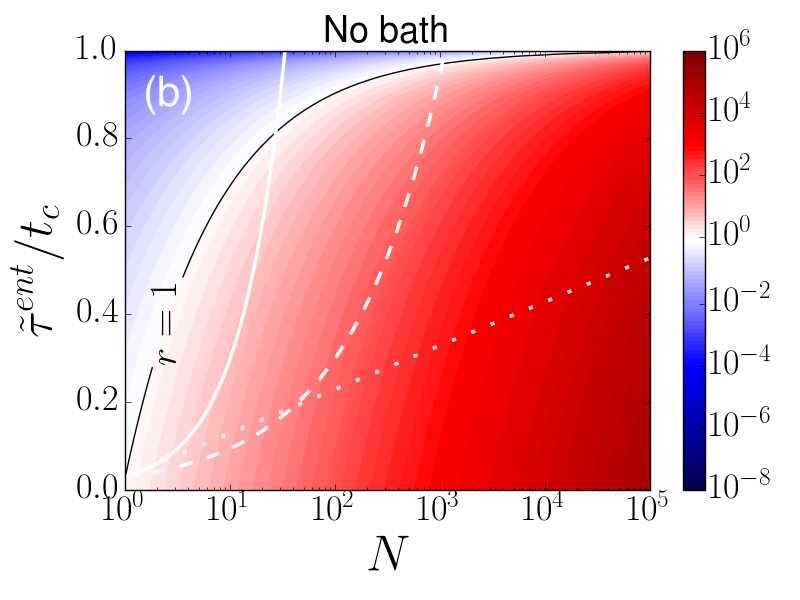}}
    \href{https://youtu.be/CC6Ljvognio}{\includegraphics[width=59mm]{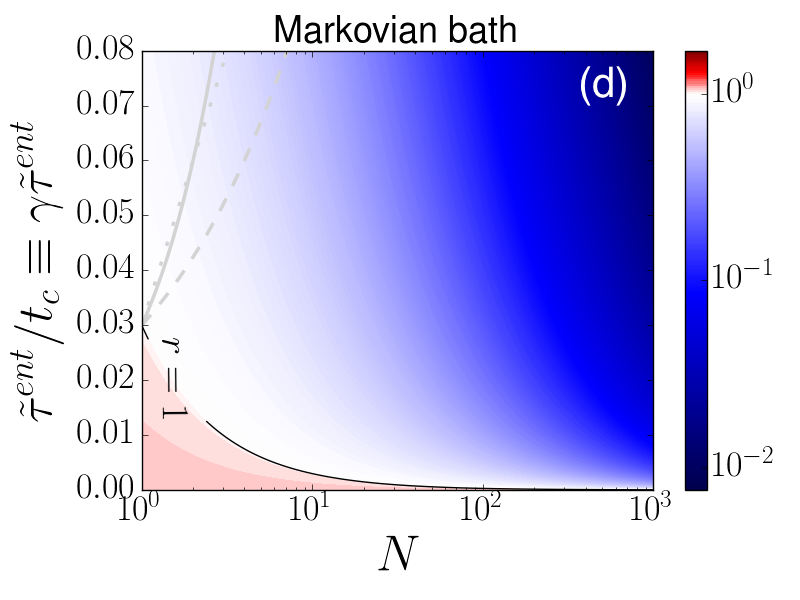}} 
    \href{https://youtu.be/ZVViwJ6pcrw}{\includegraphics[width=59mm]{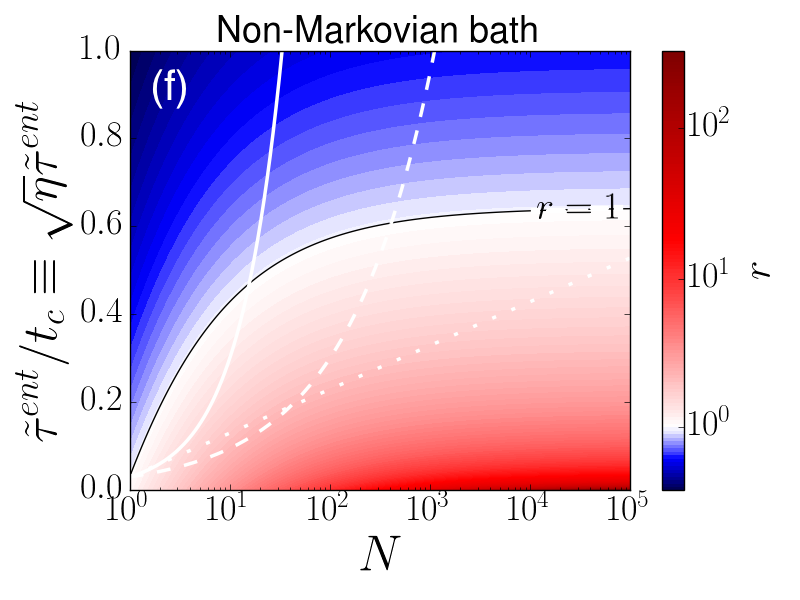}}
    \caption{Metrological gain $r$ is a function of three variables $\tilde{\tau}^\text{ent}/t_c$, $\tilde{\tau}^\text{sep}/t_c$ and $N$. The top row shows $r$ plotted against $\tilde{\tau}^\text{ent}/t_c$ and $\tilde{\tau}^\text{sep}/t_c$ with the third variable fixed at $N=10$. The bottom row shows $r$ plotted against $\tilde{\tau}^\text{ent}/t_c$ and $N$ with the third variable fixed at $\tilde{\tau}^\text{sep}/t_c = 0.03$. Animated versions of each of the plots above, with the third parameter varying in the time axis, are available by clicking on the plots above (online only).}
    \label{fig:contours}
\end{figure*}



\section{Isolated probe}\label{sec:isolated}


When the probe is isolated from its environment during the sensing period we have $\Gamma (\tau) = 0$ for all times $\tau$ so that $\hat{\rho}_\omega = \hat{U}_\omega\hat{\rho}_\text{prep}\hat{U}_\omega^{\dagger}$. In this case both \begin{equation} \frac{\mathcal{F}(\hat{\rho}_\omega^\text{sep}, \tau)}{\tilde{\tau} + \tau} = \frac{N\tau^2}{\tilde{\tau} + \tau} \end{equation} and \begin{equation} \frac{\mathcal{F}(\hat{\rho}_\omega^\text{ent}, \tau)}{\tilde{\tau} + \tau} = \frac{N^2\tau^2}{\tilde{\tau} + \tau} \end{equation} are increasing functions of $\tau$ so that the optimal sensing times $\tau_\text{opt}^\text{sep}$ and $\tau_\text{opt}^\text{ent}$ are the maximum available sensing time, limited only by the total measurement time $T$, i.e., $\tau_\text{opt}^\text{sep} = T - \tilde{\tau}^\text{sep}$ and $\tau_\text{opt}^\text{ent} = T - \tilde{\tau}^\text{ent}$. However, this is unrealistic since, in practice, the time $t$ available for each measurement is always limited by decoherence. Thus -- although the probe is isolated from the environment -- we assume that each run is limited to at most the probe system coherence time $t_c$, which gives $\tau_\text{opt}^\text{sep} = t_c - \tilde{\tau}^\text{sep}$ and $\tau_\text{opt}^\text{ent} = t_c - \tilde{\tau}^\text{ent}$. Substituting into Eq. \ref{eq:general_r} gives: \begin{equation}  r = N \left( \frac{1 - \tilde{\tau}^\text{ent}/t_c}{1 - \tilde{\tau}^\text{sep}/t_c} \right)^2 . \label{eq:r_no_noise_2} \end{equation} For the entangled state strategy to be advantageous we require $r>1$. It is straightforward to show that $r$ is a decreasing function of $\tilde{\tau}^\text{ent}/t_c$. In other words, as the entangled state preparation and readout time $\tilde{\tau}^\text{ent}/t_c$ increases (with $\tilde{\tau}^\text{sep}/t_c$ and $N$ held fixed) the metrological gain $r$ decreases. This can be seen in Fig. \ref{fig:contours}(a), where $r$ is plotted against $\tilde{\tau}^\text{ent}/t_c$ and $\tilde{\tau}^\text{sep}/t_c$ for a fixed value of $N$, and in Fig. \ref{fig:contours}(b) where $r$ is plotted against $\tilde{\tau}^\text{ent}/t_c$ and $N$ for a fixed value of $\tilde{\tau}^\text{sep}/t_c$. Since $r$ is a decreasing function of $\tilde{\tau}^\text{ent}/t_c$ we can find a critical value of $\tilde{\tau}^\text{ent}/t_c$ above which the separable state strategy outperforms the entangled state strategy. From Eq. \ref{eq:r_no_noise_2} we obtain: \begin{equation} r > 1 \implies \frac{\tilde{\tau}^\text{ent}}{t_c} < 1 - \frac{1 - \tilde{\tau}^\text{sep}/t_c}{\sqrt{N}} . \end{equation} This threshold is plotted in Figs. \ref{fig:contours}(a) and \ref{fig:contours}(b) as the black lines labelled $r=1$ that divide the red ($r>1$) and blue ($r<1$) regions.

From Eq. \ref{eq:r_no_noise_2} we see that when $\tilde{\tau}^\text{sep} = \tilde{\tau}^\text{ent} = 0$ we recover the familiar result $r=N$, indicating the Heisenberg scaling advantage. When $\tilde{\tau}^\text{sep} = \tilde{\tau}^\text{ent} \neq 0$ we still have $r = N$, though care must be taken to distinguish between \emph{absolute} Heisenberg scaling $\delta\omega_\text{opt}^\text{ent}\propto 1/N$ and \emph{relative} Heisenberg scaling $r \propto N$ (which does not necessarily imply $\delta\omega_\text{opt}^\text{ent}\propto 1/N$). Indeed, we note that Fig. \ref{fig:contours}(a) apparently shows a region of super-Heisenberg relative sensitivity ($r>N$) when $\tilde{\tau}^\text{ent}/t_c < \tilde{\tau}^\text{sep}/t_c$, although the absolute sensitivity $\delta\omega_\text{opt}^\text{ent}$ cannot exceed the Heisenberg limit.





Eq. \ref{eq:r_no_noise_2} also shows that relative Heisenberg scaling $r \propto N$ is achieved if $\tilde{\tau}^\text{ent}/t_c$ and $\tilde{\tau}^\text{sep}/t_c$ are both independent of the number of particles $N$. In general, however, the state preparation and readout times will depend on $N$. For instance, suppose that we start with the pure state $\ket{1}^{\otimes N}$, which for simplicity we assume can be generated in a negligible time. The separable state $\ket{\psi^\text{sep}}$ is straightforwardly prepared from this state by a $\frac{\pi}{2}$-rotation of each two-level system in its Bloch sphere. Often, this can be done in a time that is independent of $N$ by applying the same rotation to each of the two-level systems simultaneously. For the entangled state, the time required to prepare $\ket{\psi^\text{ent}}$ from $\ket{1}^{\otimes N}$ depends on the particular GHZ-state generation scheme. One common proposal is to evolve by the Hamiltonian $\hat{H}_\text{int} = \hbar \chi \sum_{i,j} \hat{\sigma}_x^{(i)} \otimes \hat{\sigma}_x^{(j)}$ since, after an evolution time $\tau_\text{prep}^\text{ent} = \pi/8\chi$, this leads to a GHZ state. There are some proposed implementations of $\hat{H}_\text{int}$ for which the coupling parameter $\chi$ (and hence the GHZ state preparation time $\tau_\text{prep}^\text{ent} = \pi/8\chi$) is independent of $N$, but such schemes often depend on the capabilities of the particular physical implementation, for example the intrinsic interaction of Bose-Einstein condensates \cite{You-03} or the precise tunability of parameters in superconducting circuits \cite{Wan-10}. For many other implementations (e.g. \cite{Aga-97, Mol-99, Doo-14, Doo-16a}) $\chi$ is a decreasing function of $N$ so that the preparation time $\tau_\text{prep}^\text{ent} = \pi/8\chi$ is increasing with $N$. Assuming that $\tilde{\tau}^\text{ent}/t_c$ is increasing in $N$, with $\tilde{\tau}^\text{sep}/t_c$ staying constant, we find that the entangled state strategy is always outperformed by the separable state strategy for large enough $N$, that is, there is always a finite cut-off value $N_\text{cutoff}$ such that $r < 1$ for $N > N_\text{cutoff}$. For example, in Fig. \ref{fig:contours}(b) we plot $r$ as a function of $N$ for three different types of scaling for $\tilde{\tau}^\text{ent}/t_c$: logarithmic scaling $\tilde{\tau}^\text{ent}/t_c = (1 + \log_2 N)\tilde{\tau}^\text{sep}/t_c$ (the dotted white line), square-root scaling $\tilde{\tau}^\text{ent}/t_c = \sqrt{N} \tilde{\tau}^\text{sep}/t_c$ (the dashed white line), and linear scaling $\tilde{\tau}^\text{ent}/t_c = N \tilde{\tau}^\text{sep}/t_c$ (the solid white line). In each case, the white line eventually crosses the $r=1$ threshold as $N$ increases, although the value of $N_\text{cutoff}$ can change by orders of magnitude depending on the $N$-scaling of $\tilde{\tau}^\text{ent}/t_c$. However, we note that when there is a cut-off value $N_\text{cutoff}$ there will also be a smaller number of particles $N_\text{max}$ that maximises the metrological gain $r$, i.e., $\max_N r(N) = r(N_\text{max})$. For a large number $N \gg N_\text{max}$ of two-level systems one can maintain a constant gain $\sim r(N_\text{max})$ by dividing the ensemble of $N$ systems into $\sim N/N_\text{max}$ sub-ensembles, each one prepared in a maximally entangled state of size $N_\text{max}$. A widely used figure-of-merit to assess metrological schemes is the asymptotic $N$-scaling advantage in sensitivity using entangled states. These results indicate that it may be unrealistic to expect such an advantage when state preparation and readout times are taken into account -- even for an isolated probe.




\section{Interaction with the environment} 

In the previous discussion we assumed that the system was isolated from the bath and we artificially included the effect of decoherence by simply limiting the total per-round interrogation time to $t\leq t_c$. A more realistic model is to include the effect of dephasing due to interaction with the bath. When $\Gamma (\tau) > 0$ the optimal separable state sensing time $\tau_\text{opt}^\text{sep}$ is found by solving the equation: \begin{equation} 2 \tau_\text{opt}^\text{sep} \frac{d\Gamma (\tau)}{d\tau} \Big|_{\tau = \tau_\text{opt}^\text{sep}} = 1 +  \frac{ \tilde{t}^\text{sep}}{\tilde{t}^\text{sep} + \tau_\text{opt}^\text{sep}} , \label{eq:opt_sens_sep} \end{equation} (which comes from differentiating $\mathcal{F}(\hat{\rho}_\omega^\text{sep},\tau)/(\tilde{\tau}^\text{ent} + \tau)$ with respect to $\tau$ and setting the result equal to zero). Similarly, the optimal entangled state preparation time is found by solving: \begin{equation} 2 N \tau_\text{opt}^\text{ent} \frac{d\Gamma (\tau)}{d\tau} \Big|_{\tau = \tau_\text{opt}^\text{ent}} = 1 + \frac{\tilde{t}^\text{ent}}{\tilde{t}^\text{ent} + \tau_\text{opt}^\text{ent}} . \label{eq:opt_sens_ent} \end{equation} From a microscopic model with each probe particle linearly coupled to a bath of harmonic oscillators one can derive different forms for $\Gamma (\tau)$ depending on the details of the bath (see Appendix \ref{app:bath}). Following \cite{Mat-11, Chi-12}, we consider two different limits: the Markovian limit and the non-Markovian limit.

\subsection{Markovian limit}

For a Markovian bath one can show that $\Gamma (\tau) = \gamma \tau$, with the corresponding single-spin coherence time $t_c = 1/\gamma$ (see Appendix \ref{app:bath}). These dynamics can be modelled by the master equation $\dot{\rho} = - \frac{i}{\hbar} [\hat{H}_\omega,\rho] + \frac{\gamma}{2}\sum_{i=1}^N \left( \hat{\sigma}_z^{(i)}\rho\hat{\sigma}_z^{(i)} - \rho \right) $. Estimation of $\omega$ with this form of dephasing was studied by Huelga and co-workers \cite{Hue-97} who found that $r=1$ (without taking the preparation and readout times $\tilde{\tau}^\text{sep}$ and $\tilde{\tau}^\text{ent}$ into account). Using other entangled states instead of maximally entangled GHZ states (e.g. squeezed states) can give an improvement, but only up to a constant factor $r \leq e \approx 2.7$ \cite{Esc-11}.



Taking into account the preparation and measurement times $\tilde{\tau}^\text{sep}$ and $\tilde{\tau}^\text{ent}$ we find [by solving Eqs. \ref{eq:opt_sens_sep} and \ref{eq:opt_sens_ent} for $\Gamma(\tau) = \gamma\tau$] that: \begin{equation}  \tau_\text{opt}^\text{sep} = \frac{1}{4\gamma} + \sqrt{ \left( \frac{\tilde{\tau}^\text{sep}}{2} + \frac{1}{4\gamma} \right)^2 + \frac{\tilde{\tau}^\text{sep}}{2\gamma} } - \frac{\tilde{\tau}^\text{sep}}{2} , \label{eq:Markov_t_opt_sep} \end{equation} and \begin{equation}  \tau_\text{opt}^\text{ent} =  \frac{1}{4N\gamma} + \sqrt{ \left( \frac{\tilde{\tau}^\text{ent}}{2} + \frac{1}{4N\gamma} \right)^2 + \frac{\tilde{\tau}^\text{ent}}{2N\gamma} } - \frac{\tilde{\tau}^\text{ent}}{2} . \label{eq:Markov_t_opt_ent} \end{equation} Substituting these expressions into Eq. \ref{eq:general_r} gives $r$ as a function of three variables: the entangled state preparation and readout time $\tilde{\tau}^\text{ent}/t_c = \gamma \tilde{\tau}^\text{ent}$, the separable state preparation and readout time $\tilde{\tau}^\text{sep}/t_c = \gamma \tilde{\tau}^\text{sep}$, and the number of spins $N$.


In Appendix \ref{app:dec_r} we show that, as in the case of the isolated probe, $r$ is a decreasing function of $\tilde{\tau}^\text{ent}/t_c = \gamma \tilde{\tau}^\text{ent}$ (with $\tilde{\tau}^\text{sep}/t_c$ held fixed). The question then is to find the point at which this decreasing function crosses the $r = 1$ threshold for metrological gain. By substituting $\tilde{\tau}^\text{ent} = \tilde{\tau}^\text{sep}/N$ into Eq. \ref{eq:general_r}, Eq. \ref{eq:Markov_t_opt_sep} and Eq. \ref{eq:Markov_t_opt_ent} it is straightforward to verify that this is the crossing point. The $\tilde{\tau}^\text{ent} = \tilde{\tau}^\text{sep}/N$ threshold is plotted in Figs. \ref{fig:contours}(c) and \ref{fig:contours}(d) as the black lines labelled $r=1$. When $\tilde{\tau}^\text{sep} = \tilde{\tau}^\text{ent} = 0$ this condition is satisfied and we reclaim the $r = 1$ result of Huelga and co-workers \cite{Hue-97}. A more interesting case, however, is when $\tilde{\tau}^\text{sep} \neq 0$, since the preparation and readout times are always non-zero in practice. In this case the condition $\tilde{\tau}^\text{ent} = \tilde{\tau}^\text{sep}/N$ tells us the preparation and readout of the entangled state should be a factor of $N$ times faster than for the separable state to achieve $r=1$. In an experiment this may be difficult, especially since pure separable states are often prepared as a first step towards generating entangled states. Realistically speaking we have $\tilde{\tau}^\text{ent} > \tilde{\tau}^\text{sep}/N$ which implies that the optimal precision using the entangled state that is worse than the optimal precision using the separable state, i.e. $r<1$. We again consider the three cases: logarithmic scaling, square-root scaling and linear scaling of $\tilde{\tau}^\text{ent}/t_c$. The results are plotted in Fig. \ref{fig:contours}(d) in the dotted, dashed and solid gray lines. We see that -- in each case -- as $N$ increases the entangled state strategy becomes progressively worse compared to the separable state strategy.

These results indicate that for a probe that undergoes parallel Markovian dephasing, a non-negligible preparation and readout time is more damaging to the entangled state strategy than it is to the separable state strategy. This raises questions about whether non-zero preparation and readout times might destroy the gain that is possible for non-Markovian dephasing when preparation and readout times are negligible \cite{Mat-11, Chi-12}. We now consider the non-Markovian case.








\subsection{Non-Markovian limit}

For a static or low-frequency bath one can show that $\Gamma(\tau) = \eta \tau^2$, with a corresponding single-spin coherence time $t_c = 1/\sqrt{\eta}$ (see Appendix \ref{app:bath}). Without taking measurement and preparation times into account a scaling advantage $r \propto N^{1/2} $ can be achieved in this case \cite{Mat-11, Chi-12}. Including measurement and preparation times, the optimal sensing times $\tau_\text{opt}^\text{sep}$ and $\tau_\text{opt}^\text{ent}$ are found by solving Eqs. \ref{eq:opt_sens_sep} and \ref{eq:opt_sens_ent} for $\Gamma (\tau) = \eta \tau^2$. The expressions are given in Appendix \ref{app:non-mark}. Using these solutions, $r$ is again a function of the three variables $\tilde{\tau}^\text{sep}/t_c = \sqrt{\eta}\tilde{\tau}^\text{sep}$, $\tilde{\tau}^\text{ent}/t_c = \sqrt{\eta}\tilde{\tau}^\text{ent}$ and $N$. We plot the behaviour of $r$ with respect to these variables in Figs. \ref{fig:contours}(e) and \ref{fig:contours}(f). We have numerically verified that $r$ is a decreasing function of $\tilde{\tau}^\text{ent}/t_c$ for $N$ in the range $1 - 10^{10}$ and $\tilde{\tau}^\text{sep}/t_c$ in the range $0 - 10$. We have not been able to derive an analytic expression for the $r=1$ threshold in terms of $N$ and $\tilde{\tau}^\text{sep}/t_c$, but Fig. \ref{fig:contours}(e) suggests that it is of the form $\tilde{\tau}^\text{ent}/t_c = m(N) \tilde{\tau}^\text{sep}/t_c + c(N)$ for some functions $m(N)$ and $c(N)$. The animation of Fig. \ref{fig:contours}(e) [available by clicking on the figure (online only)] shows that as $N$ increases the region of metrological gain (the red region) increases in size. This is in contrast to the Markovian case where the region of metrological gain shrinks rapidly as $N$ increases [see the animation corresponding to Fig. \ref{fig:contours}(c)]. However, this argument assumes that $\tilde{\tau}^\text{ent}$ is independent of $N$. In Fig. \ref{fig:contours}(f) the dotted, dashed and solid white lines show the trajectory of $r$ for the logarithmic, square-root and linear $N$-scalings of $\tilde{\tau}^\text{ent}$ in the non-Markovian case. It is clear that a modest precision gain is possible, but only up to a cutoff value $N < N_\text{cutoff}$ that depends on the form of the $N$-scaling of the entangled state preparation and readout time. However, as mentioned previously, in this case a constant gain can be maintained by grouping the $N$ particles into entangled sub-ensembles of the optimal size $N_\text{max}$. Finally, one can show that $\tilde{\tau}^\text{ent} = \tilde{\tau}^\text{sep}/\sqrt{N}$ implies that $r = \sqrt{N}$ [the labelled black line in Fig. \ref{fig:contours}(e)]. Since it is very difficult to prepare and readout the entangled state a factor of $\sqrt{N}$ times more quickly than the separable state, this suggests that a precision gain $r = \sqrt{N}$ would be very challenging in practice.

\section{Conclusion} 

To determine whether quantum sensors in entangled states can give a precision gain it is important to carefully consider the practical details of metrological schemes. We find that when state preparation and readout times are taken into account, a maximally entangled GHZ state can only give an advantage over a separable state if the entangled state preparation and readout time is lower than a certain threshold that depends on the number of probe particles $N$ and on the separable state preparation and readout time $\tilde{\tau}^\text{sep}$. Often, the entangled state preparation and readout times will increase with the number of particles $N$. In this case the entangled state strategy will give an advantage only up to some finite number of probe particles $N_\text{cutoff}$. The conditions to achieve a precision gain with the entangled state strategy are more difficult to achieve if dephasing is taken into account. The basic reason for the decrease in performance of the entangled state strategy in the presence of dephasing is that $\tau_\text{opt}^\text{ent}$ decreases as $N$ increases, with $\tau_\text{opt}^\text{ent} \to 0$ in the large $N$ limit, while $\tau_\text{opt}^\text{sep}$ is independent of $N$. This means that when $N$ is large there are many more prepare-sense-readout rounds in the total time $T$ for the entangled state strategy than there are for the separable state. There are thus many more preparation and readout periods that take away from the portion of $T$ that is available for sensing. We expect our conclusions to be valid for any metrological scheme in which $\tau_\text{opt}^\text{ent} \to 0$ in the large $N$ limit, for example, frequency estimation with spin squeezed states \cite{Ula-01, Tan-15}, or frequency estimation in the presence of transversal Markovian dephasing with GHZ states \cite{Cha-13} or with one-axis twisted spin-squeezed states \cite{Bra-15}. 

However, the prospects for frequency estimation with entangled probes are not completely negative, even taking preparation and readout times into account. We have shown that some gain is possible with non-Markovian dephasing and, as discussed at the end of section \ref{sec:isolated}, the metrological gain $r$ can be maintained at its maximum value by dividing the $N$ probe particles into entangled sub-ensembles. Also, depending on the specifics of the entangled state generation and readout scheme, small improvements in $r$ should be possible by preparing a partially entangled state instead of a maximally entangled GHZ state. Such states may be more robust to decoherence than the GHZ state so that their optimal sensing times would be longer and, moreover, it is likely that partially entangled states can be prepared and measured more quickly than GHZ states. Finally, there are various methods that may enhance the metrological gain that can be achieved using the GHZ state such as quantum error correction \cite{Arr-14, Kes-14, Dur-14, Lu-15}, adaptive feedback schemes \cite{Yua-15, Sek-16a, Sek-16, Liu-16}, or fast preparation \cite{Mic-03, Lap-12, Wan-10, Tan-15} and readout of entangled states. We leave investigation of these as future work.



\section{Acknowledgements}

We are very grateful to Yuichiro Matsuzaki for helpful comments. We acknowledge support from the MEXT KAKENHI Grant No. 15H05870.

\bibliographystyle{apsrev4-1} 
\bibliography{/Users/dooleysh/Dropbox/physics/BibTexLibrary/refs}

\appendix

\section{Microscopic model for the bath}\label{app:bath}

We assume that each two-level system interacts with an independent thermal bath of harmonic oscillators. The total Hamiltonian is $\sum_{i=1}^N \hat{H}^{(i)}$ where the Hamiltonian for the $i$'th two-level system and $i$'th bath is \cite{Leg-87, Sch-07}: \begin{eqnarray} \hat{H}^{(i)} &=& \frac{\hbar\omega}{2}\hat{\sigma}_z^{(i)} + \hbar \sum_k \Omega_k \hat{a}_{k}^{(i) \dagger} \hat{a}_{k}^{(i)} \nonumber\\ && + \hbar \hat{\sigma}_z^{(i)} \otimes \sum_k \left(g_k \hat{a}_{k}^{(i) \dagger} + g_k^* \hat{a}_{k}^{(i)} \right) . \end{eqnarray} Choosing a continuous Ohmic spectral density of the form $J(\Omega) = \sum_k |g_k|^2 \delta(\Omega - \Omega_k) = 4 \alpha \Omega e^{-\Omega/\Omega_c}$ where $\Omega_c$ is a cutoff frequency and $\alpha$ is a dimensionless constant, and assuming that $\beta\Omega_c \gg 1$ where $\beta$ is the inverse temperature of the bath gives a decay exponent \cite{Sch-07}: \begin{equation} \Gamma (\tau) = \frac{\alpha}{2}\ln (1 + \Omega_c^2 \tau^2 ) + \alpha \ln \left[ \frac{\sinh (\pi \tau /\beta)}{\pi \tau / \beta} \right] . \label{eq:Gamma} \end{equation} If $\tau \gg \beta$ then the second term of Eq. \ref{eq:Gamma} dominates and we have \cite{Sch-07} $\Gamma (\tau) \approx \gamma \tau$ where $\gamma = \alpha\pi/\beta$. The condition $\tau \gg \beta$ can be satisfied either by long times $\tau$ or by high temperature $1/\beta$. The form $\Gamma(\tau) \propto \tau$ is also obtained by an alternative analysis where one makes a Markovian assumption for the bath \cite{Sch-07}. For this reason $\Gamma (\tau) = \gamma \tau$ is called the Markovian limit.

When $\tau \ll \beta$ and $\Omega_c \tau \ll 1$ the first term of Eq. \ref{eq:Gamma} dominates. Approximating $\ln (1 + \Omega_c^2 \tau^2 ) \approx \Omega_c^2 \tau^2$ gives $\Gamma (\tau) \approx \eta \tau^2$ where $\eta = \alpha \Omega_c^2 / 2$. Since $\Gamma (\tau) \not\propto \tau$ we call these dynamics non-Markovian. The condition $\tau \ll \beta$ is satisfied for short times or for low bath temperature, while the condition $\Omega_c \tau \ll 1$ is satisfied for short times or for interaction with primarily low-frequency bath modes (i.e. low cutoff frequency $\Omega_c$).

\section{Optimal sensing times in the non-Markovian case}\label{app:non-mark}

For the non-Markovian case, the optimal sensing times $\tau_\text{opt}^\text{sep}$ and $\tau_\text{opt}^\text{ent}$ are found by solving Eq. \ref{eq:opt_sens_sep} and Eq. \ref{eq:opt_sens_ent} in the main text, with $\Gamma(\tau) = \eta \tau^2$. The resulting solutions are: \begin{eqnarray} \tau_\text{opt}^\text{sep} &=&  \frac{z^\text{sep}}{3} - \frac{1}{3z^\text{sep}} \left(\tilde{\tau}^{\text{sep}^2} + \frac{3}{4\eta}\right) - \frac{\tilde{\tau}^\text{sep}}{3} , \\  \tau_\text{opt}^\text{ent} &=& \frac{z^\text{ent}}{3} - \frac{1}{3z^\text{ent}} \left(\tilde{\tau}^{\text{ent}^2} + \frac{3}{4N\eta}\right) - \frac{\tilde{\tau}^\text{ent}}{3} , \end{eqnarray} where \begin{eqnarray}  z^\text{sep} &=& e^{i4\pi/3} \Bigg[ \tilde{\tau}^{\text{sep}^3} - \frac{45\tilde{\tau}^\text{sep}}{8\eta} + \nonumber \\ && \sqrt{\left( \tilde{\tau}^{\text{sep}^3} - \frac{45\tilde{\tau}^\text{sep}}{8\eta} \right)^2 - \left( \tilde{\tau}^{\text{sep}^2} + \frac{3}{4\eta} \right)^3} \Bigg]^{1/3} , \\ z^\text{ent} &=& e^{i4\pi/3}\Bigg[ \tilde{\tau}^{\text{ent}^3} - \frac{45\tilde{\tau}^\text{ent}}{8N\eta} + \nonumber \\ && \sqrt{\left( \tilde{\tau}^{\text{ent}^3} - \frac{45\tilde{\tau}^\text{ent}}{8N\eta} \right)^2 - \left( \tilde{\tau}^{\text{ent}^2} + \frac{3}{4N\eta} \right)^3} \Bigg]^{1/3} . \end{eqnarray}


\section{Outline of a proof that $r$ decreases with $\tilde{\tau}^\text{ent}/t_c$ in Markovian case}\label{app:dec_r}

We would like to show that $r$ is a decreasing function of $\tilde{\tau}^\text{ent}/t_c \equiv \gamma \tilde{\tau}^\text{ent}$ in the Markovian case when $\Gamma(\tau) = \gamma \tau$. For clarity, we denote the variable of interest as $x = \tilde{\tau}^\text{ent}/t_c$ and we also write the expression for $r$ as $r(x) = c r_1(x) r_2 (x)$ where $c  = N (\gamma\tilde{\tau}^\text{sep} + \gamma \tau_\text{opt}^\text{sep})e^{2\gamma\tau_\text{opt}^\text{sep}}/(\gamma\tau_\text{opt}^\text{sep})^2$ is a non-negative number that is independent of $x$ and $r_1(x) = \gamma\tau_\text{opt}^\text{ent}(x) / (x + \gamma \tau_\text{opt}^\text{ent}(x))$ and $r_2(x) = \gamma\tau_\text{opt}^\text{ent}(x) e^{-2N\gamma\tau_\text{opt}^\text{ent}(x)} $ are both non-negative functions of $x$. Here \begin{equation} \gamma \tau_\text{opt}^\text{ent}(x) = \frac{1}{4N} + \sqrt{ \left( \frac{x}{2} + \frac{1}{4N} \right)^2 + \frac{x}{2N} } - \frac{x}{2} , \end{equation} is easily shown to be an increasing function of $x$. If we can show that $r_1(x)$ and $r_2(x)$ are both decreasing functions, then this would prove that $r$ is a decreasing function because $\frac{dr}{dx} = c \frac{dr_1}{dx}r_2 + cr_1 \frac{dr_2}{dx} \leq 0$ (since $c, r_1, r_2 \geq 0$). We see that $r_2(x)$ is a decreasing function since it is the composition $r_2(x) = f(\tau_\text{opt}^\text{ent}(x))$ of the increasing function $\tau_\text{opt}^\text{ent}(x)$ and the function $f(y) = \gamma y e^{-2N\gamma y}$, which is decreasing for $y \geq 1/2\gamma N$. Since $\tau_\text{opt}^\text{ent}(x) \geq 1/2\gamma N$ for any $x$ we see that $r_2$ is decreasing.

To show that $r_1(x)$ is a decreasing function we observe that it can be written in the form $r_1(x) = \frac{1 - g(x)}{1 + g(x)}$ where $g(x) = \frac{2Nx}{1 + \sqrt{(2Nx + 1)^2 + 8Nx}}$. Differentiating $r_1$ with respect to $x$ gives $\frac{dr_1}{dx} = -\frac{dg(x)/dx}{1 + g(x)}\left[ 1 + \frac{1 - g(x)}{1 + g(x)} \right]$. Since $1 \pm g(x) > 0$ we see that $r_1$ is a decreasing function of $x$ if $\frac{dg(x)}{dx} \geq 0$. We find that \begin{eqnarray} \frac{dg(x)}{dx} &=& \frac{2N}{1 + \sqrt{(2Nx + 1)^2 + 8Nx}} \times \\ && \left[ 1 - \frac{4N^2x^2 + 6Nx}{4N^2x^2 + 12Nx + 1 + \sqrt{(2Nx + 1)^2 + 8Nx}} \right] , \nonumber\end{eqnarray} which is always non-negative (since the term in square brackets is always non-negative).

\end{document}